\documentclass[11pt,a4paper,numbers]{article}
\pdfoutput=1
\usepackage{jcappub}

\usepackage{amssymb,mathtools,amsmath,latexsym,graphics, graphicx,epsfig,multirow,comment,hyperref,appendix, verbatim} 
\usepackage{pifont}
\usepackage{lipsum}

\usepackage{slashed}
\usepackage{subfigure}
\usepackage{feynmp}
\usepackage{graphicx}
\usepackage{amsfonts}
\usepackage{color}
\usepackage{cancel}

\def\hbar{\hspace{0pt}\raisebox{1pt}{$-$} \hspace{-7pt} h}

\def\5{\overline 5}

\definecolor{JJ}{RGB}{0,144,255}

\newcommand{\be}{\begin{equation}}
\newcommand{\ee}{\end{equation}}
\newcommand{\bea}{\begin{eqnarray}}
\newcommand{\eea}{\end{eqnarray}}

\newcommand{\ba}{\begin{eqnarray}}
\newcommand{\ea}{\end{eqnarray}}

\begin{document}
\title{Halo-Independent Direct Detection of Momentum-Dependent Dark Matter}

\author{John F. Cherry,$^{1}$}

\author{Mads T. Frandsen,$^{2}$}

\author{and Ian M. Shoemaker$^{2}$}

\affiliation{$^{1}$Theoretical Division, Los Alamos National Laboratory, Los Alamos, New Mexico 87545, USA}
\affiliation{$^{2}$CP$^{3}$-Origins and the Danish Institute for Advanced Study, University of Southern Denmark, Campusvej 55, DK-5230 Odense M, Denmark}

\arxivnumber{LA-UR-14-23073, CP3-Origins-2014-018 DNRF90,DIAS-2014-18}
\abstract{

We show that the momentum dependence of dark matter interactions with nuclei can be probed in direct detection experiments without knowledge of the dark matter velocity distribution. This is one of the few properties of DM microphysics that can be determined with direct detection alone,  given a signal of dark matter in multiple direct detection experiments with different targets. Long-range interactions arising from the exchange of a light mediator are one example of momentum-dependent DM. For data produced from the exchange of a massless mediator we find for example that the mediator mass can be constrained to be $\lesssim 10$ MeV for DM in the 20-1000 GeV range in a halo-independent manner. 
}



\maketitle


\section{Introduction}

A definitive, non-gravitational detection of Dark Matter (DM) has yet to occur.  However a large number of low-background, direct searches for DM are underway with the goal of detecting the feeble nuclear recoil { energy} { deposited by DM particles passing through the detector}. The main target of these experiments are Weakly-Interacting Massive Particles (WIMPs), { which are the most thoroughly studied DM candidates}. { The attractiveness of WIMP Dark Matter is} driven by the fact that the relic {  abundance is controlled by their annihilation cross section}.  In the Early Universe, WIMPs are kept in thermal equilibrium by number-changing interactions, $\overline{X}X \leftrightarrow \overline{f}f$, where $f$ is some SM particle.  Eventually though, as the Universe expands and WIMPs are diluted, these number-changing interactions cease, and the abundance of WIMPs ``freezes out." 
Given that this paradigm requires DM to share some interactions with the SM, it provides many experimental lines of inquiry, and assuming that DM interacts with quarks or gluons it can be probed at direct detection experiments. 

{ In view of the null results from direct detection and the LHC, simple models of thermal relic WIMPs termed Effective Field theory (EFT) models or `Maverick' \cite{Beltran:2010ww} models where the DM particle itself is the only new particle accessible at LHC,  are nearly ruled out~\cite{Goodman:2010yf,Bai:2010hh,Fox:2011pm}. }
{ However, this conclusion is easily evaded when the EFT approach itself is not valid}, as in the case of a mediator much lighter than the DM. Then annihilation of DM to a pair of mediators typically dominates over other available annihilation channels. In the case of asymmetric DM, the annihilation cross section requirement is even more { stringent because} one needs to ``annihilate away'' the symmetric abundance.  
{ Thus a light mediator coupling to DM remains a viable venue for symmetric or asymmetric DM}. 

In contrast with high-energy colliders, direct detection offers a sensitive probe of such light mediators { at the cost of relying on the galactic DM halo to provide collisional energy.  As a consequence of this} direct detection experiments suffer from some uncertainty in the astrophysical distribution of DM.  { To combat this uncertainty}, direct detection analysis using astrophysics-independent methods for interpreting data has gained increased interest recently \cite{Drees:2008bv,Fox:2010bz, Frandsen:2011gi,DelNobile:2013cva,Feldstein:2014gza,Fox:2014kua}. 

In the present paper, we illustrate the utility of these methods in determining the spin-independent of the DM scattering  { and outline a new method which is also agnostic with respect to the velocity distribution of DM in the halo}. {For simplicity we restrict this study to elastic, spin-independent scattering from single component DM.}  Using this simple method, we perform a projection study of what up-coming ton-scale experiments can say about the presence of light mediators, and more generally momentum dependence of the scattering cross section. {Similar projections have been recently made for momentum-independent cross sections ~\cite{Pato:2010zk,Pato:2012fw} and momentum-dependent cross sections~\cite{McDermott:2011hx} { assuming specific models of the DM velocity distribution}. 
}

\section{Direct Detection in $v_{\rm min}$-space}

Direct detection involves a combination of dark matter particle physics, { nuclear physics} and astrophysics. It has been pointed out that in the case of { simple} spin-independent interactions, one can ``integrate out'' the DM astrophysics and compare experiments { without any assumptions about the unknown local DM distribution~\cite{Fox:2010bu,Fox:2010bz}. }
These methods have since been extended to cover momentum-dependent~\cite{DelNobile:2013cva} and inelastic scattering~\cite{Bozorgnia:2013hsa,Scopel:2014kba}. 


For elastic,  spin-independent scattering the differential scattering rate (not yet including detector effects) at a direct detection experiment is given by 
\be \frac{dR}{dE_{R}} = \sum_j \mathcal{F}_j\frac{dR_j}{dE_{R}} = \sum_j \mathcal{F}_j\frac{1}{2\mu_{nX}^{2}} \left[ \frac{f_{p}}{f_{n}} Z_j + (A_j-Z_j)\right]^{2} F^{2}_j (E_{R}) \tilde{g}(v_{{\rm min},j}),
\ee
where 
\be
\tilde{g}(v_{{\rm min},j}) = \frac{ \rho \sigma_{n}}{m_{X}} \int_{v_{{\rm min},j}(E_{R})}^{\infty} \frac{f({\bf v}+{\bf v}_{E}(t))}{v} d^{3} v, 
\ee
with $\sigma_{n}$ the DM-neutron scattering cross section, $m_{X}$ is the dark matter mass, the index $j$ denotes the individual types of nuclide present in the detector medium with mass number $A_j$ and proton number $Z_j$.  $\mathcal F_j$ is the number fraction of nuclide $j$, $F_j (E_{R})$ is the nuclear form factor, $\mu_{nX}$ is the nucleon-DM reduced mass,  $f_{p}/f_{n}$ is the ratio of the { effective proton and neutron couplings to DM \cite{Jungman:1995df}.}\footnote{Though we do not consider ``isospin-violating'' couplings in this paper, we note that both next-to-leading order effects~\cite{Cirigliano:2012pq,Cirigliano:2013zta} and hadronic uncertainties~\cite{Crivellin:2013ipa} can be sizeable.} Finally $f({\bf v}+{\bf v}_{E}(t))$ is the local DM velocity distribution evaluated in the galactic rest frame and ${\bf v}_{E}(t)$ is the velocity of the Earth relative to the galactic rest frame. 

The minimum velocity for an incoming DM particle to produce a nuclear recoil of energy $E_{R}$ is
\be
v_{{\rm min},j} (E_{R})= \sqrt{m_{N}E_{R}/2\mu_j^{2}} \, ,
\label{eq:vminoEr}
\ee
with $\mu_j$ being the DM-nuclide reduced mass, and $m_{N}$ is the mass of the nuclide.  Lastly we also have in the above the astrophysics parameters $\rho$ and $f(v)$ describing the local DM mass density and velocity distribution respectively. There is considerable uncertainty on these astrophysics parameters, and the DM inferences (e.g. its mass and scattering cross section) one {na\"ively} draws from direct detection depend sensitively on their values.

Rather than assuming the values of $\rho$ and $f(v)$, we can instead report halo-independent constraints or parameter estimations on the quantities $m_{X}$, $v_{\rm min}$, and $\tilde{g}(v_{\rm min})$. Thus, assuming that $v_{{\rm min},j}$ is independent of the nuclide, for each fixed DM mass $m_{X}$ we can simply use the observed rate $dR/dE_{R}$ to infer $\tilde{g}(v_{\rm min})$ as a function of $v_{\rm min}$:
\be 
\tilde{g}(v_{\rm min}) = \sum_j \frac{2 \mu_{nX}^{2}}{\mathcal F_j \left[f_{p}/f_{n} Z_j + (A_j-Z_j)\right]^{2}F_j^{2}(E_{R})} \frac{dR_j}{dE_{R}}.
\label{eq:money}
\ee
In performing such a mapping to $v_{\rm min}$-space in this way, we have assumed that the only momentum dependence in the scattering is in the nuclear form factor, where the relation between momentum transfer $q$ and recoil energy is given by $q=\sqrt{2 m_N E_R}$.  We also assume that the experimenter has no ability to distinguish between recoils produced by different nuclides within the detector medium.

\section{Momentum-Dependent Dark Matter Scattering in $v_{\rm min}$-space: \\ the $\tilde{g}$ ratio test}

\label{sec:main}

Let us now illustrate how a straightforward extension of this method can be used to extract information about the momentum dependence from direct detection data without making assumptions about the DM astrophysics. {Here we aim for a qualitative discussion of the method and delay the introduction of detector effects such as finite energy resolution until Sec.~\ref{sec4}.}

Above in Eq.(\ref{eq:money}) we assumed that the energy-dependence from the DM microphysics is trivial, i.e. all the energy dependence is encoded in the nuclear form factor $F(E_{R})$. There are many ways in which this assumption can be violated. An especially simple example, is that DM may exchange a very light mediator with nuclei.  In this case, the cross section will scale as $d\sigma/dE_{R} \propto q^{-4}$. A simple parameterization~\cite{Chang:2009yt} of non-trivial spin-independent is 
\be \frac{d \sigma}{dE_{R}} =  \left(\frac{d \sigma}{dE_{R}}  \right)_{0} \left(\frac{q^{2}}{q_{\rm ref}^{2}}\right)^{n} \left(\frac{q_{\rm ref}^{2}+m_\phi^{2}}{q^{2}+m_\phi^{2}}\right)^{2},
\label{eq:sigmamom}
\ee
where 
$m_{\phi}$ is the mass of the exchanged mediator, and we fix $q_{\rm ref} = 10$ MeV throughout.  Above, $\left(d \sigma/dE_{R}  \right)_{0}$ is the standard spin-independent DM-nucleus cross section
\be 
\left(\frac{d \sigma}{d E_{R}}\right)_{0} = \frac{m_{N}}{2\mu_{nX}^{2}v^{2}} \tilde{\sigma}_{n} F^{2} (E_{R}).
\label{Eq:qparamsimple}
\ee
Written in this way, the integer $n$ parameterizes the unknown Lorentz structure of the DM-quark operator, while the momentum dependence from the propagator is explicitly factored out.~\footnote{We note that there are many interactions not described by Eq.~(\ref{eq:sigmamom}) For example, DM magnetic dipole moment scattering contain multiple terns with differing momentum dependences. Moreover DM bound-state scattering or break-up~\cite{Laha:2013gva} introduce additional sources of momentum dependence not encompassed by~Eq.(\ref{eq:sigmamom}).}  Thus the standard spin-independent contact interaction scattering corresponds to the $n=0$, heavy mediator limit $m_{\phi}^{2} \gg q^{2}$.  The momentum transfer $q$ is related to $v_{\rm min}$ as $q = 2\mu v_{\rm min}$.  Note that the quantity $\tilde{\sigma}_{n}$ is equal to the standard spin-independent cross section on neutrons only for $n=0$ contact interactions. More generally the quantity $\tilde{\sigma}_{n}$ is a parameter with units of cross section which is a function of $m_\phi$, $q_{\rm ref}$, and $n$.  The function of $\tilde{\sigma}_{n}$ is to fix the normalization of the rate for more complicated momentum-dependent interactions.

\begin{figure}[t!] 
\begin{center}
 \includegraphics[width=.47\textwidth]{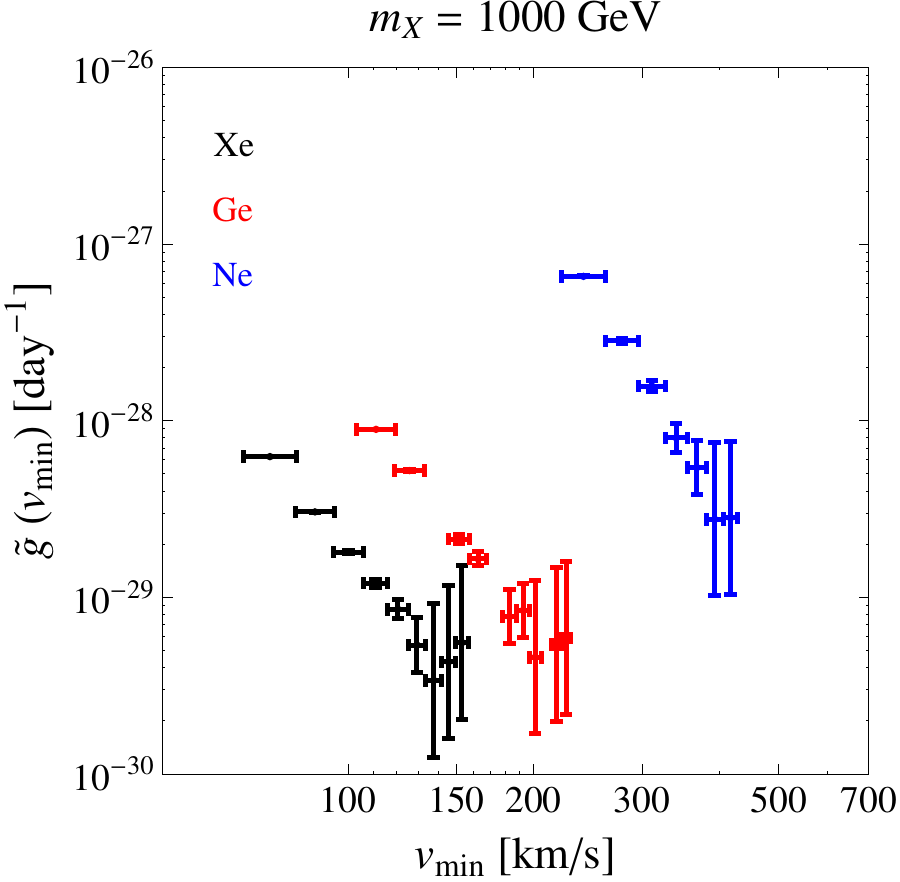}
 \includegraphics[width=.45\textwidth]{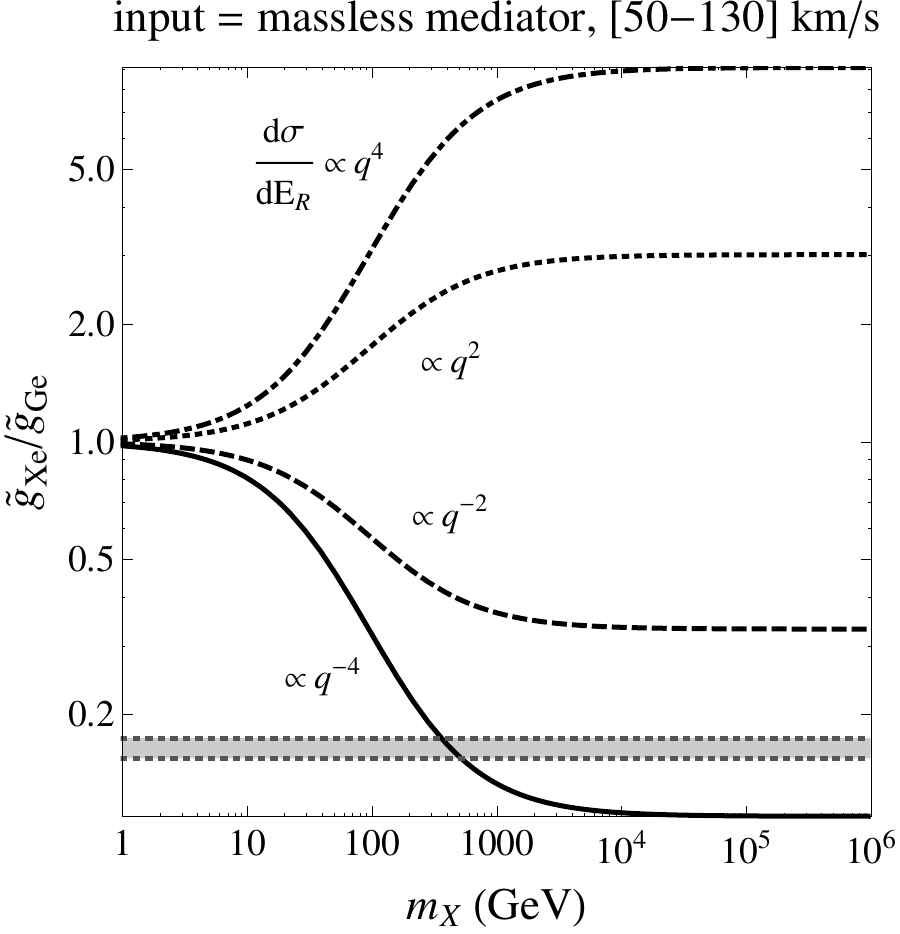}

\caption{{\it Left}: Here the input data (from scattering via a massless mediator) is incorrectly mapped (assuming momentum-independent scattering) into $(\tilde{g}-v_{\rm min})$ space using Eq.~(\ref{eq:money}). However where data bins overlap, the $\tilde{g}$ ratios yield information about the momentum-dependence.  {\it Right}: Expected $\tilde{g}$ ratio between a Xe and Ge target in which DM scattering has a non-trivial spin-independent, parameterized by the integer $n$ as shown in Eq.~(\ref{eq:sigmamom}). 
The observed $\tilde{g}$ ratio in a $[50-130]$ km/s bin is the shaded band.}
\label{fig:discrepancy}
\end{center}
\end{figure}

With the simple cross section parameterization in Eq.~(\ref{eq:sigmamom}) we are at first interested in distinguishing two possibilities: the presence or absence of momentum dependence in the DM interaction.  
{To answer this question, we can simply use Eq.~(\ref{eq:sigmamom}) with $m_{\phi} \gg q, q_{\rm ref}$, such that the cross section becomes}
\be
 \frac{d \sigma}{dE_{R}} =  \left(\frac{d \sigma}{dE_{R}}  \right)_{0} \left(\frac{q^{2}}{q_{\rm ref}^{2}}\right)^{n}.
 \label{eq:ncross}
 \ee
 In this case the signal of non-trivial momentum dependence corresponds to $n \neq 0$.  
 
 
 We now imagine the fortunate situation where two experiments employing different target materials observe a signal. If the dark matter microphysics is momentum-dependent, the data will lead to discrepant inferences of $\tilde{g}(v_{\rm min})$ under the assumption of Eq.~(\ref{eq:money}). 
Suppose experiment $i$ uses a target nucleus of mass $m_{N_{i}}$, $i=1,2$. {With the signal spectrum produced from scattering of the type in Eq.~(\ref{eq:ncross}), we expect experiment $i$ to infer a value for $\tilde{g}(v_{\rm min})$ of}
\be 
\tilde{g}_{\rm infer,i} = \tilde{g}_{\rm true}\left(\frac{q^{2}}{q_{\rm ref}^{2}}\right)^{n} = \tilde{g}_{\rm true} \left(\frac{2\mu_{i}v_{\rm min}}{q_{\rm ref}}\right)^{2n}\ee 
%
where $\mu_{i} = m_{X}m_{N_i}/(m_{X} + m_{N_i})$. Taking the ratio of the inferred value of $\tilde{g}(v_{\rm min})$ at the same $v_{\rm min}$ by two different experiments we obtain (assuming spin-independent scattering~\footnote{An analogous expression for spin-dependent scattering can be written down after constructing the $\tilde{g}_{ \rm{infer}}$ using the spin-dependent form factor and average spin contributions from the proton and neutron spin groups.}):
\be 
\frac{\tilde{g}_{\rm infer,1} }{\tilde{g}_{\rm infer,2} }\biggr\rvert_{v_{\rm min}} = \left(\frac{\mu_{1}}{\mu_{2}}\right)^{2n} .\label{eq:simpgtilderatio}
\ee

At DM masses $m_X \gg m_{N_i}$ where $\mu_{i} \simeq m_{N_{i}}$, we expect a significant discrepancy to appear, implying momentum dependence of the true scattering cross-section. Instead at low DM masses  $m_X \ll m_{N_i}$ when $\mu_{i} \simeq m_{X}$ and the momentum transfer is nearly identical for different targets, we expect little difference in the inferred values of $\tilde{g}$ for different experiments and it will not be possible to distinguish between the above scenarios, e.g. \cite{Frandsen:2013cna}. 

We illustrate this in the left panel of Fig.~\ref{fig:discrepancy}, where we plot the ratio of the inferred values of $\tilde{g}(v_{\rm min})$ { in  Xenon, Germanium, and Neon experiments, using $m_{X} = 1000\, \rm GeV$, $m_\phi = 0$, Eq.~(\ref{eq:sigmamom}),} and assuming Helm nuclear form factors.  Thus at each value of $v_{\rm min}$ that is compared, one expects an offset in the inferred value of $\tilde{g}$. {The right hand panel of Fig.~\ref{fig:discrepancy} illustrates the DM mass dependence of this offset, confirming that so long as we are not in the regime where $m_{X} \ll m_{N_i}$ momentum dependence will produce an offset that can be employed as a diagnostic of the momentum dependence of the scattering.}

{This simple $\tilde{g}$ ratio measure is easily extended to include the mediator mass dependence by using Eq.~(\ref{eq:sigmamom}) such that Eq.(\ref{eq:simpgtilderatio}) generalizes to}
\be 
\frac{\tilde{g}_{\rm infer,1} }{\tilde{g}_{\rm infer,2} }\biggr\rvert_{v_{\rm min}} = \left(\frac{\mu_{1}}{\mu_{2}}\right)^{2n} \left(\frac{4\mu_{2}^{2}+\left(m_\phi / v_{\rm min}\right)^{2}}{4\mu_{1}^{2}+\left(m_\phi / v_{\rm min}\right)^{2}}\right)^{2}.
\label{eq:gtilderatio}
\ee

Our strategy for determining the momentum dependence of the dark matter scattering is therefore to map the experimental data on $dR/dE_R$ into bins in $v_{\rm min}$ space and compare overlapping bins.  {In general, this mapping will be different for distinct experiments with different target nuclei and isotopic abundances, and will also depend on the hypothetical DM mass which we wish to test.  This provides the benefit that our uncertainties on the velocity distribution of the DM in the galactic halo are guaranteed to cancel away when taking a ratio of two independent measurements.  So long as the DM is single-component, the distinct experiments are guaranteed to sample precisely the same portion of the $\tilde g\left(v_{\rm min}\right)$ curve.  This provides a measurement of the DM interaction physics regardless of the de-facto shape of $\tilde g\left(v_{\rm min}\right)$, provided that $m_X$ is not so light that the DM-Nucleus reduced masses of all experiments are degenerate.}  In the case where no bins overlap we can, in principle, extrapolate a best fit curve of $\tilde{g}\left( {v_{\rm min}}\right)$ through the available bins and then compare the inferred $\tilde{g}\left( {v_{\rm min}}\right)$ functions. However this re-introduces dependence on the velocity function {by way of forcing an observer to assume that the velocity distribution of the DM in the galactic halo corresponds to a known class of analytic functions (e.g. the Standard Halo Model) and the associated observational uncertainties.} 

Since we are interested in extracting the dependence of the cross-section on the momentum transfer, or equivalently the recoil energy, it is important to know the momentum dependence inherent in the nucleon and nuclear form factors.  A general momentum dependence of the spin-independent differential dark matter-nucleus scattering on a given nuclide per unit time and detector mass, assuming a single DM species, may be written as \cite{Cirigliano:2012pq}
\begin{multline*} 
\frac{dR}{dE_{R}} =  \tag{3.7}  \\
 \frac{\kappa_X \rho}{m_X} \int_0^\infty \left[\sum_i  ( f_{p}^i (q,v) F^i_p (q,v) Z + (A-Z)f_{n}^i(q,v) ) F^i_n (q,v) - T_2(q,v) \right]^2 \frac{f({\bf v}+{\bf v}_{E}(t))}{v} d^{3}v,
\end{multline*}
where $\kappa_X$ is a factor specific to the nature of the DM particle, $i$ runs over the exchanged mediators and we include both separate nuclear form factors $F^i_{n,p} (q,v)$ for protons and neutrons and a form factor for two nucleon interactions $T_2(q,v)$ \cite{Cirigliano:2012pq,Cirigliano:2013zta} in addition to the single nucleon form factors $f_{p,n}^i (q,v)$. We also imagine a separation of $f^i_p (q)=f^i_{p,X} (q) f^i_{p,N} (q)$ where the first factor gives the particle physics momentum dependence of the interaction, { arising from the DM-mediator vertex} while the second factor gives the hadronic physics form factor. 
Thus in our example above in Eq.~(\ref{eq:ncross}) we have assumed $f^i_{p,X} (q)= f^i_{n,X} (q) \sim q^{2n}$, $f^i_{n,N} (q)=f^i_{n,N}$, $T_2(q,v)=0$ and an appropriate nuclear form factor $F^i_{n} (q,v)=F^i_{p} (q,v)=F(E_R)$.

For a scalar mediator, interacting with DM and quarks via Yukawa couplings with zero tree-level momentum dependence, the momentum-dependent part of  $f_{n,p} (q)$ and $T_2(q,v)$ gives corrections to the rate of the order of a few percent below 100 keV nuclear recoil energies when the heavy quark couplings are suppressed, but can be large when unsuppressed~\cite{Cirigliano:2012pq,Cirigliano:2013zta}.  The same is true for the momentum dependence of the Helm form factor $F(E_R)$ and more importantly we expect differences between the Helm form factor (included in the analysis) and a more { accurate} form factor are not above this level.  Extracting the momentum dependence from DM particle physics is therefore feasible.

We provide the details of our analysis below, {and a complete description of our techniques in Appendix~\ref{appendix}.}



\section{Detector Mock-ups and Input Spectrum}
\label{sec4}
\begin{table}[t]
\centering
\fontsize{9}{9}\selectfont
\begin{tabular}{c|ccc}
\hline
\hline
 target & $\epsilon_{eff}$ [ton$\times$yr] & $E_{thr}$ [keV] & $\sigma(E)$ [keV]  \\
\hline
Xe & 0.88 & 5 & $0.6~ {\rm keV} \sqrt{E_{R}/{\rm keV}} $\\
Ge & 0.88 & 5 & $\sqrt{(0.3)^{2}+(0.06)^{2}E_{R}/{\rm keV}} \,\rm keV$\\
Ne & 0.88 & 5 &$1~ {\rm keV} \sqrt{E_{R}/{\rm keV}}$\\

\hline
\end{tabular}
\caption{\fontsize{9}{9}\selectfont Characteristics of future direct dark matter experiments using Xenon, Germanium and Neon as target nuclei. Here $\epsilon_{eff}$ is the effective exposure, $E_{thr}$ is the low-energy threshold, and $\sigma(E)$ is the energy resolution of the experiment.}\label{tabExp}
\end{table}

To reasonably well simulate the near-term experimental capabilities, we include efficiencies, energy resolution, exposures, and background expectations. Similar theoretical projections have been made previously with momentum-independent~\cite{Pato:2010zk,Pato:2012fw} and momentum-dependent cross sections~\cite{McDermott:2011hx} { assuming specific models of the DM velocity distribution}. {We use 5 keV energy thresholds for Xenon, Germanium, and Neon targets. Given that large-scale xenon and germanium experiments with $< 5$ keV thresholds already exist, the adoption of 5 keV should be conservative. Although achieving such a low threshold for neon may be an experimental challenge, the results of the present paper motivate a low-mass target, low-threshold experiment.}

{For simplicity we shall assume Gaussian energy resolution. Though we note that this is not always the case in real detectors, we expect the correction to be small.} The Gaussian energy resolution for each target is specified in Table~\ref{tabExp}.  
{ The number of events expected in the energy range $[E_1,E_2]$ is:
\begin{equation}
N(E_1, E_2) = {\rm Exp} \, \int {\rm Res} (E_1, E_2, E_R) \, \epsilon \frac{dR}{dE_R} dE_R\, \;,
\label{eq:Nevents}
\end{equation}
where $\epsilon$ is the efficiency, $\rm Exp$ is the raw exposure (not including detector/analysis cut efficiencies), and ${\rm Res} (E_1, E_2, E_R)$ is the detector response function taken to be }
\be
{\rm Res} (E_1, E_2, E_R) =\frac{1}{2} \left[ {\rm erf} \left(\frac{E_{2}-E_{R}}{\sqrt{2} \sigma(E_{R})} \right) - {\rm erf} \left(\frac{E_{1}-E_{R}}{\sqrt{2} \sigma(E_{R})} \right) \right].
\ee


In this work we assume a raw exposure of $2.2$ ton-years and a flat efficiency, $\epsilon = 0.4$, yielding an effective exposure of 0.88 ton-years unless otherwise stated. Lastly, we assume that each mock experiment will achieve their stated goals of reaching a $<1$ background event expectation.

Though the velocity distribution remains unknown it has become canonical to assume a Maxwell-Boltzmann (MB) distribution~\cite{Drukier:1986tm,Freese:1987wu}, truncated at the escape speed. We use this standard halo model (SHM) as our fiducial $f(v)$ choice to generate mock experimental spectra and note that detailed simulations of DM structure formation do not appear to deviate markedly from this choice~\cite{Kuhlen:2009vh}. 

In the rest frame of the Milky Way the velocity distribution,	
\begin{equation}
  f_{MB}(\vec{v}) = \left\{
     \begin{array}{lr}
       N  e^{-v^2/v_{0}^{2}}, & v < v_{esc}\\
       0, & v > v_{esc}
     \end{array}
   \right. ,
   \label{gauss_param}
\end{equation}
is determined by its dispersion $v_{0}$ and the local escape speed $v_{esc}$. In this case the velocity integral $g(v_{\rm min})$ corresponding to the SHM has a closed form expression \cite{Smith:1988kw,Jungman:1995df,Savage:2006qr,McCabe:2010zh}.

In the SHM the dispersion is equated to the circular speed which is observable and measured to be $\langle v_{e}(t) \rangle = 230$ km/s~\cite{Bovy:2009dr,McMillan:2009yr}. We use a local DM density $\rho_{DM} = 0.3~{\rm GeV}{\rm cm}^{-3}$ and escape speed $v_{esc} = 550$ km/s.


\section{Results}

To generate our mock signals, we have taken representative points in parameter space consistent with the LUX null results, specifically the $90\%$ confidence limit upper bound on the DM-nucleon cross section~\cite{Akerib:2013tjd}. The details of our LUX treatment can be found in~\cite{Cirigliano:2013zta,Frandsen:2013bfa,Frandsen:2014ima}.  {Fits to the individual data sets for each detector medium are performed with standard Binned Maximum Likelihood (ML) analysis~\cite{PDG:2013xk}, which in addition to evaluating the goodness of fit for $n$, $m_\phi$, and $m_X$ optimizes the best fit in the velocity dispersion, $v_{0}$, of the SHM.  The individual datasets are then combined into a master dataset where the same ML analysis is performed for a \lq\lq global\rq\rq\ analysis.  {The details of this analysis can be found in Appendix \ref{app:a1}.} Once these tests are complete, the $\tilde g$ ratio test is also conducted on the original data set for every $v_{\rm min}$ bin in which multiple experiments share observed events.  Because we are exploring a three dimensional parameter space, we present the confidence interval data after it has been marginalized over one of the parameters (listed on the top of the figure with the true value of the parameter in the legend).  When performing the marginalization procedure, we assume a flat prior for the entire parameter space.}

\begin{figure*}[t!] 
\begin{center}
      \includegraphics[width=.45\textwidth]{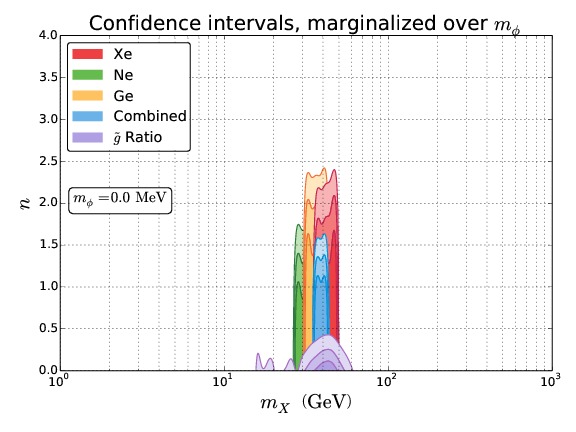}
  \includegraphics[width=.45\textwidth]{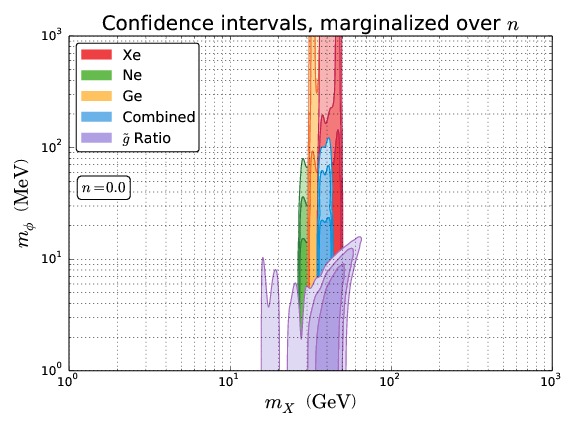}
    \includegraphics[width=.45\textwidth]{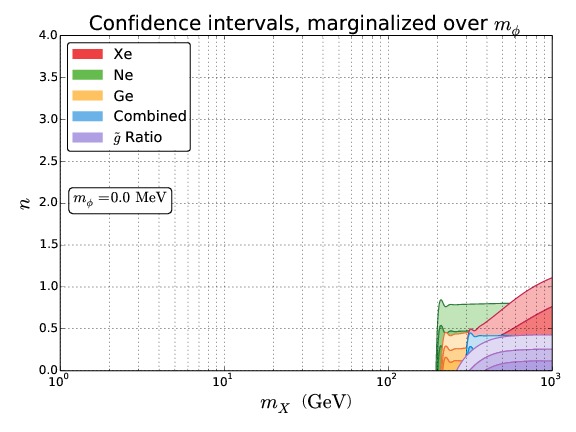}
  \includegraphics[width=.45\textwidth]{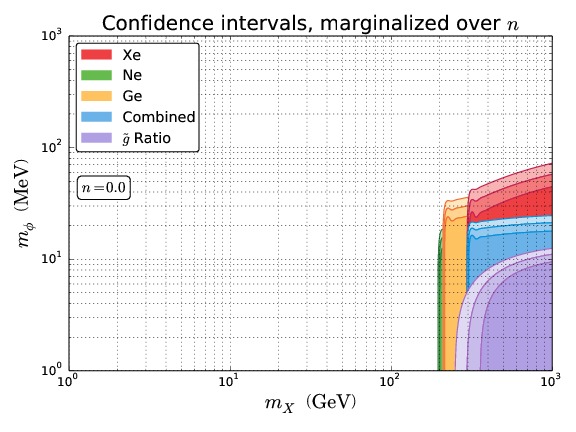}

\caption{Reconstruction of DM parameters from an input spectrum generated via a massless mediator. The results on the first row correspond to DM mass $m_{X} =  20$ GeV, with $n=0$, $\tilde\sigma_n = 3.8\times 10^{-43}\, \rm cm^{2}$ and total event counts for each detector medium $\{ N_{Ne} = 186, N_{Ge} = 123, N_{Xe} = 143\}$.  The bottom row shows results for $m_X = 1000 $ GeV, with $\tilde\sigma_n = 5.7\times 10^{-42}\, \rm cm^{2}$ and total event counts for each detector medium $\{ N_{Ne} = 178, N_{Ge} = 119, N_{Xe} = 87\}$.  For each 2D plot the three parameters $(n, m_{X}, m_{\phi})$ are fit to the mock data with one of the three parameters marginalized over.  Figures on the left have been marginalized over $m_\phi$ and figures on the right have been marginalized over $n$. For each data set the 1, 2 and 3 $\sigma$ confidence levels are shown from darkest to lightest shading respectively. }
\label{fig:mass40}
\end{center}
\end{figure*}


As a first example, let us examine the results coming from the $m_{X} = 40$ GeV case with a massless mediator with $n=0$. This would arise from the exchange of a light ({$m_\phi \ll q$}) vector or scalar.  The marginalised $(1,2,3)\sigma$ best-fit contours are displayed in both the $(m_{\phi},m_{X})$ and $(n,m_{X})$ planes in the upper panel of Fig.~\ref{fig:mass40}. There we see that in addition to the benefit of having multiple targets with signal data, we also observe the utility of the $\tilde{g}$ ratio test as defined in Sec.~\ref{sec:main}. {In this example, the mediator mass is limited to a smaller upper bound of $m_{\phi} \lesssim 10$ MeV, while the operator integer is constrained to be $n < 0.2$, and the DM mass bounded to be $m_{X} = 40^{+10}_{-8}$ GeV. In this conservative example, we have shown that one can learn the DM mass while simultaneously learning that a new light force carrier connects DM and nuclei with an identifiable momentum dependence.}

It is useful to know how general this conclusion is, and in particular if it holds for heavier DM. We simulate the situation of $10^{3}$ GeV DM again interacting with a massless mediator and display the results in the bottom panel of Fig.~\ref{fig:mass40}.  Here only a lower bound on the DM mass  $m_{X} \gtrsim 370$ GeV is possible. This is a result of a well-known $m_{X} - \sigma_{n}$ degeneracy at high DM mass that has been previously observed in spin-independent scattering (see e.g.~\cite{Pato:2010zk,Friedland:2012fa}), and arises simply from the fact that $v_{\rm min}$ becomes independent of the DM mass at high $m_{X}$.  Despite this degeneracy however both the mediator mass and operator integer remain well-constrained, $m_{\phi}\lesssim 90$ MeV, and  $n \lesssim 0.2$ respectively. 

\begin{figure*}[t!] 
\begin{center}
  \includegraphics[width=.45\textwidth]{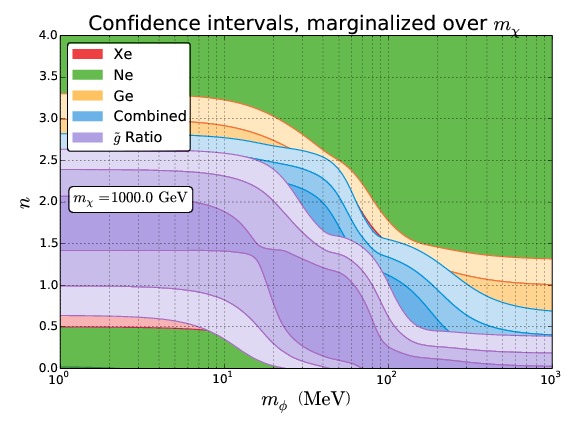}
  \includegraphics[width=.45\textwidth]{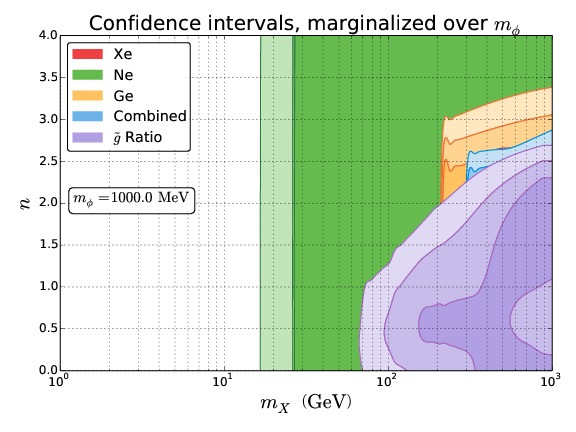}
  \includegraphics[width=.45\textwidth]{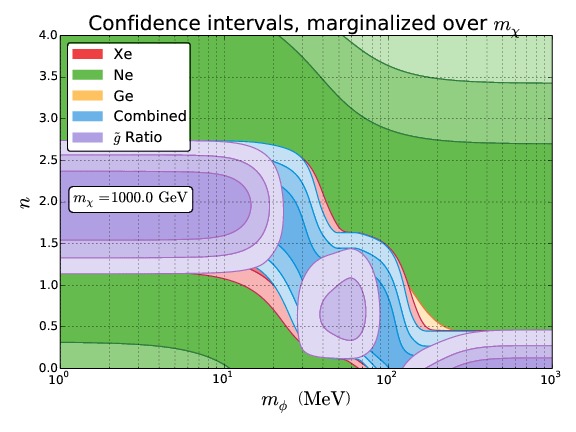}
  \includegraphics[width=.45\textwidth]{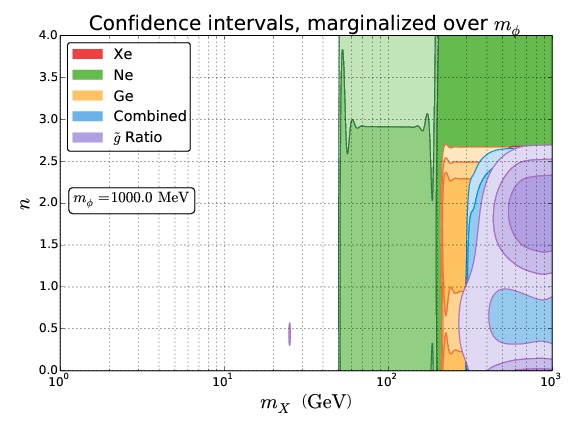}
  \includegraphics[width=.45\textwidth]{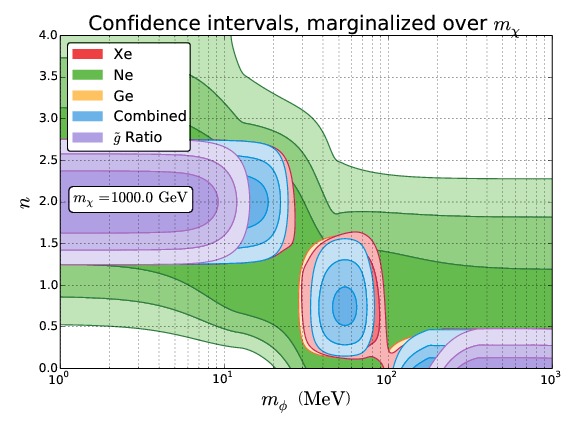}
  \includegraphics[width=.45\textwidth]{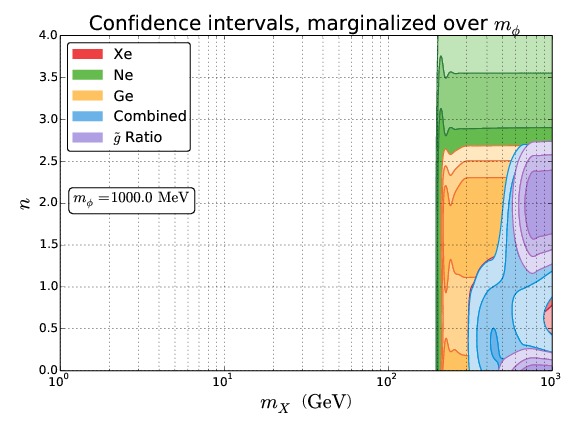}
   
\caption{{Reconstruction of DM parameters from an input spectrum generated via a contact interaction mediator.  All results correspond to DM mass $m_{X} =  1000$ GeV, with $n=0$, $m_\phi = 1000\, \rm MeV$, and $\tilde\sigma_n = 1.2\times 10^{-44}\, \rm cm^{2}$.  The total exposure time has been grouped by rows, increasing from top to bottom.  The top row shows the confidence intervals for ${\rm Exp} = 0.22 \, \rm ton\times yr$ and total event counts for each detector medium $\{ N_{Ne} = 3, N_{Ge} = 39, N_{Xe} = 63\}$.  The middle row shows the confidence intervals for ${\rm Exp} = 0.88 \, \rm ton\times yr$ and total event counts for each detector medium $\{ N_{Ne} = 10, N_{Ge} = 194, N_{Xe} = 276\}$.  The bottom row shows the confidence intervals for ${\rm Exp} = 3.52 \, \rm ton\times yr$ and total event counts for each detector medium $\{ N_{Ne} = 43, N_{Ge} = 636, N_{Xe} = 1022\}$. Figures on the left have been marginalized over $m_X$ and figures on the right have been marginalized over $m_\phi$. For each data set the 1, 2 and 3 $\sigma$ confidence levels are shown from darkest to lightest shading respectively. }}
\label{fig:mass1000}
\end{center}
\end{figure*}

\begin{figure*}[t!] 
\begin{center}
  \includegraphics[width=.45\textwidth]{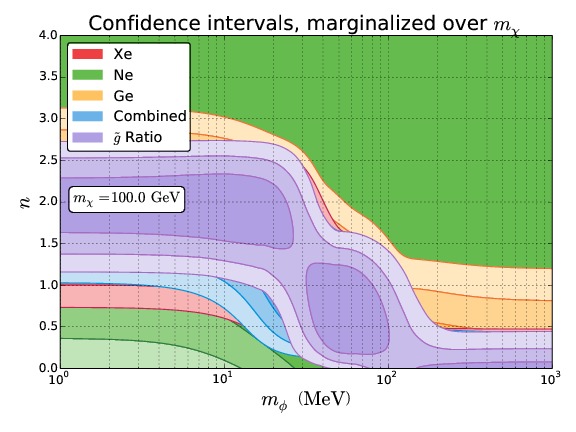}
  \includegraphics[width=.45\textwidth]{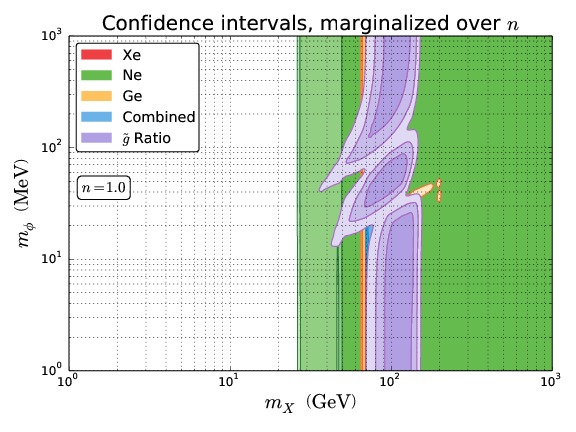}
  \includegraphics[width=.45\textwidth]{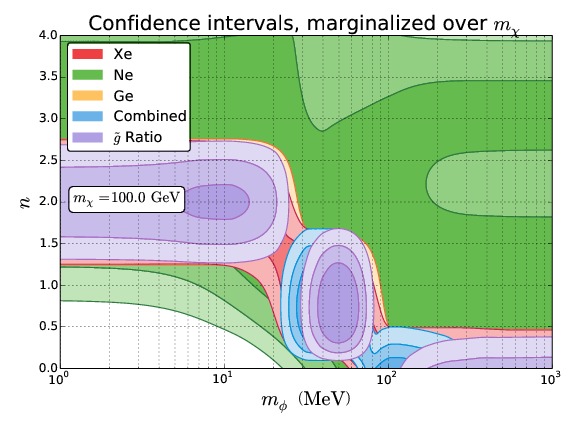}
  \includegraphics[width=.45\textwidth]{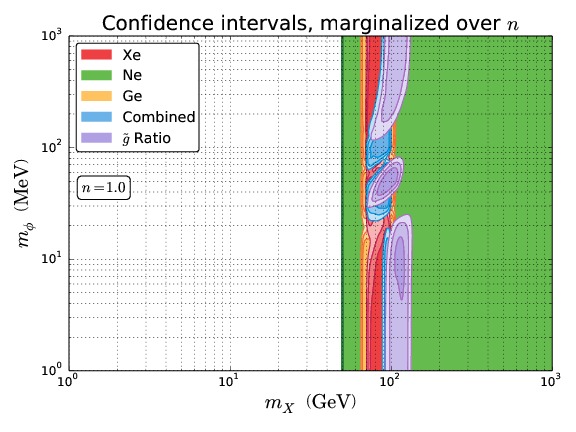}
  \includegraphics[width=.45\textwidth]{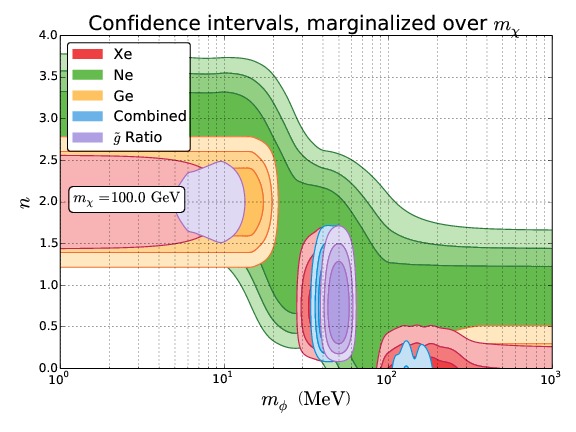}
  \includegraphics[width=.45\textwidth]{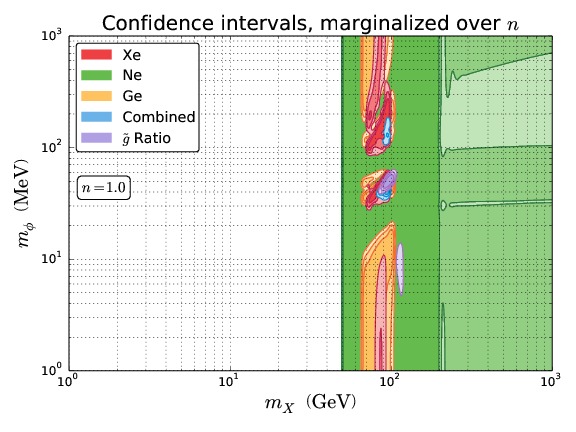}
   
\caption{{Reconstruction of DM parameters from an input spectrum generated via a intermediate mass mediator.  All results correspond to DM mass $m_{X} =  100$ GeV, with $n=1$, $m_\phi = 50\, \rm MeV$, and $\tilde\sigma_n = 2.7\times 10^{-46}\, \rm cm^{2}$.  The total exposure time has been grouped by rows, increasing from top to bottom.  The top row shows the confidence intervals for ${\rm Exp} = 0.88 \, \rm ton\times yr$ and total event counts for each detector medium $\{ N_{Ne} = 9, N_{Ge} = 163, N_{Xe} = 254\}$.  The middle row shows the confidence intervals for ${\rm Exp} = 3.52 \, \rm ton\times yr$ and total event counts for each detector medium $\{ N_{Ne} = 35, N_{Ge} = 650, N_{Xe} = 1017\}$.  The bottom row shows the confidence intervals for ${\rm Exp} = 14.08 \, \rm ton\times yr$ and total event counts for each detector medium $\{ N_{Ne} = 143, N_{Ge} = 2601, N_{Xe} = 4073\}$. Figures on the left have been marginalized over $m_X$ and figures on the right have been marginalized over $m_\phi$. For each data set the 1, 2 and 3 $\sigma$ confidence levels are shown from darkest to lightest shading respectively. }}
\label{fig:mass100}
\end{center}
\end{figure*}

{As an example of possible degeneracies with our technique, we consider the possibility of inferring that DM-nucleon scattering contains no momentum dependence. An example of this sort with 100 GeV DM is shown in Fig.~\ref{fig:mass1000}. Here we have chosen $n=0$ and $m_{\phi} = 1$ GeV such that no relevant momentum dependence enters into the scattering cross section. There we observe that both the mass of the mediator and the operator integer $n$ are not well-determined by the data. This degeneracy is linked to the phenomenon that a momentum-independent cross section could arise ``accidentally'' from the exchange of a light mediator with $n=2$, c.f Eq.~(\ref{eq:sigmamom}), such that the two types of spin-independent cancel each other out.  In this case the DM mass can be reliably reconstructed while the operator which produces the DM-quark interaction is limited to a pair-wise constraint (either $n=0$ contact interaction or $n=2$ with a massless mediator).  Also shown in Figure~\ref{fig:mass1000} is the effect of increased effective exposure for the detectors.  An additional degeneracy is present when there is insufficient data to rule out minor deviations in the inferred shape of $\tilde g\left( v_{\rm min}\right)$.  In such cases, a point in parameter space which differs by $n=1$ from the actual value is also possible.  In the example shown here, $n=1$ with $m_\phi \approx 45\, \rm MeV$ is also a degenerate possibility, but one can see in Fig.~\ref{fig:mass100} that this intermediate degeneracy is lifted with a sufficient number of events.  If the DM-nucleon interaction is mediated by either a very light (or massless) or very massive mediator, one would need orthogonal data (e.g. collider or astrophysical) to infer which class of DM microphysics is truly responsible for the interaction. 
}  

{Similar degeneracies in the inferred DM interaction parameters are present when intermediate mass mediators are considered.  Figure~\ref{fig:mass100} shows an example where $n=1$, $m_{\phi} = 50\,\rm MeV$, and $m_X = 100\,\rm GeV$, so that typically $q \sim m_\phi$ for all experiments.  The top panels of Fig.~\ref{fig:mass100} exhibit similar degeneracies to those in Fig.~\ref{fig:mass1000}.  This is due to the similarity (within errors) of the overall shape of the $\tilde g\left( v_{\rm min}\right)$ curve over the range where the experiments are capable of detecting recoils from DM interactions.  The range of this degeneracy is limited to $n\pm 1$, which corresponds to a massless mediator ($n+1$) or a contact interaction ($n-1$). However, as can be seen in the bottom panels of Fig.~\ref{fig:mass100}, with a large number of observed events the statistical errors shrink to an extent which allows the degeneracy to be completely lifted.}



\begin{figure*}[t!] 
\begin{center}
      \includegraphics[width=.45\textwidth]{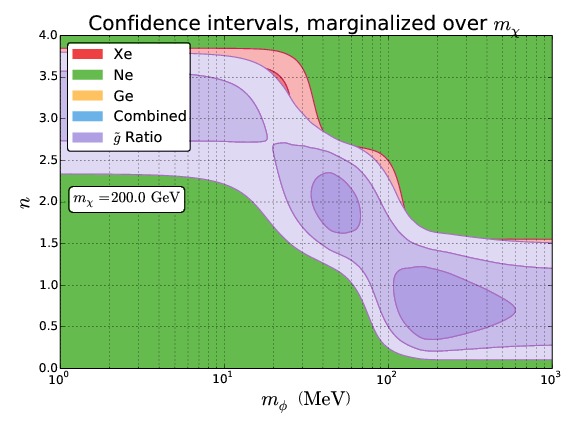}
  \includegraphics[width=.45\textwidth]{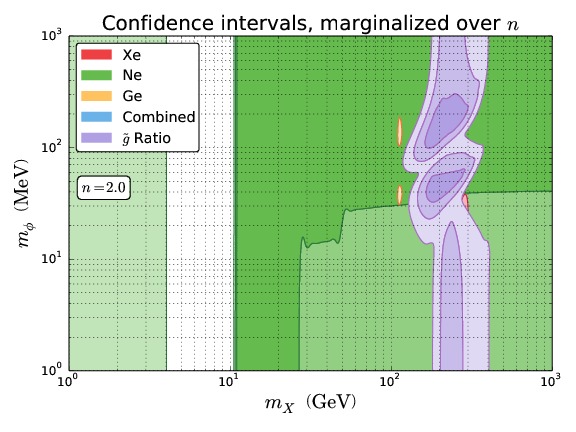}
    \includegraphics[width=.45\textwidth]{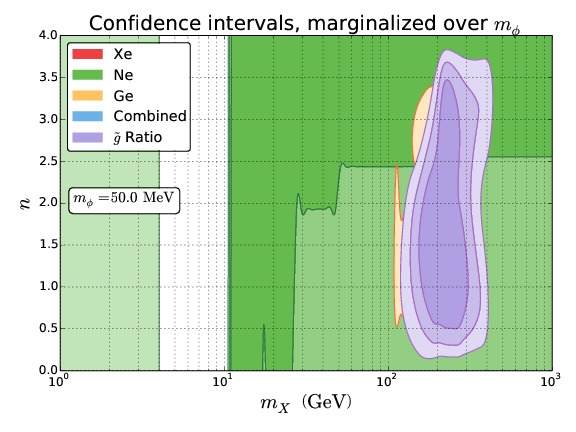}

\caption{{Reconstruction of DM parameters from an input velocity distribution which consisted of only 10 streams of dark matter.  The velocity distribution was selected randomly via Monte Carlo from a probability distribution which matches the standard SHM velocity profile with $\langle v_{e}(t) \rangle = 230$ km/s, $v_{esc} = 550$ km/s, $v_{0} = 230$ km/s.  The ML analysis was conducted under the assumption of a continuous SHM velocity distribution which has the same parameters.  The results where generated using the dark matter interaction parameters  $n=2$, $m_X = 200\,\rm GeV$, and $m_\phi = 50\,\rm MeV$, with $\tilde\sigma_n = 1.0\times 10^{-47}\, \rm cm^{2}$ and total event counts for each detector medium $\{ N_{Ne} = 1, N_{Ge} = 145, N_{Xe} = 333\}$.  Note that because of the small number of streams which dominate the dark matter velocity distribution, the ML analysis {\it rules out} the entire DM interaction parameter space with a confidence of $> 3\sigma$ for the combined fit. For each data set the 1, 2 and 3 $\sigma$ confidence levels are shown from darkest to lightest shading respectively. }}
\label{fig:MB}
\end{center}
\end{figure*}

To illustrate the true halo-independent character of the ratio test we show an example where the {ML test and the ratio test were conducted for a DM velocity distribution which consisted of only 10 individual streams of DM.  This is intended to mimic the intriguing possibility that the local DM distribution around Earth is dominated by several large streams as a consequence of the hierarchical structure formation of the Milky Way halo.  The velocity profile was constructed from 10 streams which where randomly chosen via Monte Carlo from the probability distribution function of streams within the SHM.  The prescription for finding a Monte Carlo probability distribution for the SHM is quite simple.  We exploit the fact that the Maxwell-Boltzmann velocity distribution is a three dimensional Gaussian distribution in momentum space.  This allows one to randomly generate dark matter stream momenta from the Gaussian distribution (which is cutoff above a certain momentum which corresponds to $v_{esc}$ for a given dark matter mass) and convert those momenta to stream velocities.  One then simply adds the velocity of the Earth about the galactic center to all of the stream velocity vectors and takes the resultant stream velocity magnitudes.  We also assign a weight to each stream assuming a flat prior, as not all streams in the Milky Way halo are expected to contain the same quantity of dark matter.  The sum of all weighted dark matter streams is then renormalized so that Monte Carlo realization is consistent with the LUX constraints on $\tilde g\left( v_{\rm min}\right)$.

The SHM velocity profile used for the Monte Carlo realization followed the same parameters used throughout this paper, with $\langle v_{e}(t) \rangle = 230$ km/s, $v_{esc} = 550$ km/s, $v_{0} = 230$ km/s.  An initial data set was prepared using $n=2$, $m_X = 200\,\rm GeV$, and $m_\phi = 50\,\rm MeV$ using the 10 stream velocity distribution function.}  {When performing the {ML} test, the goodness of fit was evaluated by optimizing the fit of the {\it continuous} SHM velocity profile to the observed data.  The stochastic distribution of dark matter stream velocities results in isolated experiments being able to obtain good individual fits to the original microphysics of the DM, but the combined analysis of all observations rules out the entire interaction parameter space for DM with a confidence of $> 3\sigma$.  In contrast, because the ratio test requires no assumption be made about the DM velocity distribution it is able to correctly identify the true DM interaction parameters as the best fit, along with the expected degeneracies between $n$ and $m_\phi$ for a momentum-dependent cross section discussed above.}

Lastly it is important to note that many models of DM interactions yield terms in the cross section with disparate momentum dependence. 
However even in the case that multiple terms with different $q$ dependences enter, the method illustrated presently would yield a determination of some non-integer value of $n$. Beyond such a determination, we leave the extension of this method to more general spin-independent for future work. 


\section{Conclusion}

We have presented and illustrated a new method for determining the momentum dependence of DM from direct detection data in a genuinely halo-independent manner.  In many cases the momentum dependence of the DM-nucleus scattering cross section can be well-determined from future direct detection data. The complementarity of targets aids significantly in this determination. Moreover, it is worth stressing again that this is one of the few properties of dark matter microphysics that can be extracted from near-term experiments in a manner that is independent of the DM velocity distribution. This does not {necessarily} hold for the velocity-dependence of the scattering, which {may be} degenerate with the form of the velocity distribution and {require different techniques to} be robustly determined. 

As we have argued, one especially interesting application of these methods is their ability to reveal the existence of a new light force carrier between DM and nuclei. Such mediators have been invoked both in the context of evading collider and direct detection limits while retaining a sufficiently large annihilation cross section to be a thermal relic. Moreover such mediators could be relevant for the self-scattering of DM~\cite{Feng:2009hw,Loeb:2010gj,Tulin:2013teo}.

\section*{Acknowledgements}
We would like to thank Andreas Crivellin, Eugenio Del Nobile, Martin Hoferichter, Ranjan Laha, and Stefano Scopel for useful comments. The CP3-Origins centre is
partially funded by the Danish National Research Foundation, grant number DNRF90.
MTF acknowledges a Sapere Aude Grant no. 11-120829 from the Danish Council for
Independent Research.  This work was also supported in part  by the U.C. Office of the President in conjunction with the LDRD Program at LANL.

\appendix

%
%
%
%
%

\section{Signal Analysis}
{
Now that we have established the basic ingredients which create a dark matter signal in our suite of next generation detectors, we must endeavor to determine what sort of information might be extracted from such a signal.  We are primarily interested in extracting constraints on the the momentum dependence $n$ of the dark matter interaction with ordinary nuclear matter.  It is also possible that one might infer limits on the dark matter mass and the mass of the mediator which couples DM and quarks from the halo-independent parameterization of $\tilde g\left( v_{\rm min}\right)$, in addition to the momentum dependence of the interaction.
}
\label{appendix}
{
\subsection{Binned Maximum Likelihood}
}
\label{app:a1}
{
In order to compute the expected number of events for each experiment we must specify a velocity distribution.  We follow the notation of~\cite{Savage:2006qr} for the SHM where the dispersion, escape speed, and boost velocity from the Galactic to Earth frame are respectively $v_{0}$, $v_{esc}$, and $v_{obs}$.  We will refer to the synthetic observed data as $\{ x_i\}$, which is the set of counts $x$ in bin $i$ of a given experiment (all $x_i$ are natural numbers).  Once the simulated signal has been generated for a given detector (outlined in Section~\ref{sec4}) our task turns to the business of constraining what range of dark matter interaction parameters produce a good fit to the data.  We choose to vary the momentum dependence, $n$, dark matter mass $m_X$, and the interaction mediator mass $m_\phi$ over the parameter space.}

{To evaluate the goodness of fit for each choice of parameters we generate a new set of synthetic data $\{\mu_i\}$, which is the set of expected events $\mu$ in bin $i$ of an experiment under the assumption that the point in parameter space we have selected is the {\it correct} one.  Once we have chosen a given trio of $n$ , $m_X$, and $m_\phi$ and generated our expected $\{\mu_i\}$, this allows us to normalize the overall dark matter cross section, $\sigma_X$, by scaling the predicted number of events, $\sum_i\{\mu_i\}$, to match total number of events in the simulated data set, $\sum_i\{x_i\}$.  
The normalization of the dark matter cross section is performed independently for each of the three detector media.  Our intent is to mimic the actions of independently operated detection experiments, which would reasonably choose to first analyze the data in their own detector as an isolated set of events.
}

{Having generated an expected signal $\{\mu_i\}$ we then compare the simulated signal data $\{ x_i\}$ for a given detector to the null hypothesis for the expected signal.  Explicitly we evaluate the probability that the observed data $\{ x_i\}$ is a statistical fluctuation on the expected event count $\{\mu_i\}$ for the fixed parameters $n$, $m_X$, and $m_\phi$.  This probability is then interpreted as our goodness of fit with which a given point in our parameter space matches the observed data and used to define the confidence intervals shown in the figures in the preceding sections.} 

{The general prescription for this analysis is known as the Binned Maximum Likelihood method~\cite{PDG:2013xk}.  We begin with the Poisson probability, $P_i$, for each bin in the data to match the expected number of events,
\be
P_i = \frac{\mu_i^{x_i}}{x_i !}e^{-\mu_i}\,.
\label{poiss}
\ee
The likelihood, $L$, of observing the signal $\{ x_i\}$ with expectation $\{\mu_i\}$ is given by the product of the Poisson probabilities,
\be
L\left( \{ x_i\},\{\mu_i\}\right) = \prod_i P_i = \prod_i \frac{\mu_i^{x_i}}{x_i !}e^{-\mu_i}\,.
\label{likelihood}
\ee
The probability distribution of $L$ for different sets of $\{ x_i\}$ is invariant when Eq.~\eqref{likelihood} is multiplied by an overall constant.  For reasons which will become clear shortly, we will choose to multiply Eq.~\eqref{likelihood} by $L^{-1} \left( \{ x_i\},\{x_i\} \right)$.  This will allow us to make the convenient definition of the likelihood ratio,
\be
\mathcal{L}\left( \{ x_i\},\{\mu_i\}\right) = - \ln \left[ \frac{ L\left( \{ x_i\},\{\mu_i \}\right) }{ L \left( \{ x_i \},\{x_i\} \right) } \right] \, .
\label{ratio}
\ee
With a bit of algebra it can be shown that the likelihood ratio has a simple form,
\be
\mathcal{L}\left( \{ x_i\},\{\mu_i\}\right) = \sum_i \mu_i - x_i + x_i \ln \frac{x_i}{\mu_i}\, ,
\label{LLratio}
\ee
which is much simpler to evaluate than Eqs.~\eqref{likelihood} or~\eqref{ratio} and has the benefit of being a positive number for all realizations of the expectation $\{\mu_i\}$.  An important property of $\mathcal{L}$ is that in the limit of \lq\lq large\rq\rq $\{ x_i\}$ the likelihood ratio probability distribution asymptotes to the $\chi^2$ probability distribution, with the relation that $\mathcal{L} \simeq \chi^2/2$, a result known as Wilks' Theorem~\cite{Wilks:1938uq}.}

{The goodness of fit for the point in parameter space we are considering can be determined by comparing $\mathcal{L}\left( \{ x_i\},\{\mu_i\}\right)$ to cumulative probability distribution function of $\mathcal{L}$ to compute the probability that the observed events satisfy the null hypothesis.  Generally, there is no closed analytic form for the cumulative probability distribution function of $\mathcal{L}$ (although one might use the closed form of the $\chi^2$ distribution if there are a sufficiently large number of events for Wilk's Theorem to apply), so the distribution function must be constructed via Monte Carlo methods.}  

{We begin by computing $\{ x^\prime_i\}$, a Monte Carlo realization of the expected data set $\{\mu_i\}$ constructed using the Poisson probability in Eq.~\eqref{poiss}.  For this new data set, we compute $\mathcal{L}\left( \{ x^\prime_i\},\{\mu_i\}\right)$ and record the value in a histogram.  Repeating this procedure many times will build up the probability distribution of $\mathcal{L}$ empirically.  For this work we compute up to the $3\sigma$ confidence intervals for our synthetic data sets.  This requires $P\left( 3\sigma\right) \geq 1/\sqrt{N}$, where $N$ is the number of Monte Carlo realizations we create to compute the probability distribution function, yielding $N\geq 1.4\times 10^5$ realizations.}

{In addition to evaluating the goodness of fit for the hypothetical Ne, Ge, and Xe detectors, we also evaluate the combined goodness of fit to all of the experimental data sets.  This is done simply by repeating the steps outlined in the preceding three paragraphs for the combined data set $\{ x^{combined}_j\} = \{\{ x_i^{Ne}\} ,\{ x_i^{Ge}\} ,\{ x_i^{Xe}\}\}$, with the associated combined predicted number of events   $\{ \mu^{combined}_j\} = \{\{ \mu_i^{Ne}\} ,\{ \mu_i^{Ge}\} ,\{ \mu_i^{Xe}\}\}$.}

\subsection{Ratio Test}

{There is an additional test which provides improved diagnostics of the microphysics governing the dark matter nucleon interaction which is possible only when comparing two or more experiments which have distinct detection media.  We begin by supposing that there is only a single species of dark matter which is detected by our suite of experiments.  This is by no means a certainty, but for our purposes the introduction of multi-component dark matter would produce undue complexity which is not motivated by current observational constraints.  With single component dark matter, we can say concretely that there can be only one dark matter velocity distribution which is sampled by all of the experiments, $\tilde g_{true}\left( v_{\rm min}\right)$.  However, in the instance that the momentum dependence of the dark matter interaction is a free parameter, the reconstructed function $\tilde g_{infer}\left( v_{\rm min}\right)$ is modified by both the reduced mass of the dark matter-nucleus system and the momentum dependence of the dark matter interaction, i.e. Equation~\eqref{eq:sigmamom}.  By proceeding bin by bin and comparing the overall normalization of $\tilde g_{infer}\left( v_{\rm min}\right)$ from detectors which employ different media, we can derive a constraint on the allowed ratio of $\tilde g_{infer}\left( v_{\rm min}\right)$ between the two experiments.  One can choose to compare the ratio of ${\tilde{g}_{\rm infer,1} }/{\tilde{g}_{\rm infer,2} }$, which is shown in Equation~\eqref{eq:gtilderatio}, or to reconstruct $\tilde g_{true,1}\left( v_{\rm min}\right) / \tilde g_{true,2}\left( v_{\rm min}\right)$ under a chosen set of DM interaction parameters and verify that the ratio is equal to unity.  Both of these techniques are equivalent and together they define what we call the \lq\lq Ratio\rq\rq\ test, wherein the reconstructed $\tilde g \left( v_{\rm min}\right)$ functions from multiple distinct experiments must satisfy the requirement that all experiments observe the same velocity distribution self-consistently.}

{To begin with we take the synthetic events, $\{ x_i\}$, for each detector medium which are already binned identically in $v_{\rm min}$ for a specific choice of $m_X$ according to Equation~\eqref{eq:vminoEr}.  This organization ensures that the data sets for all of the experiments, $\{ x_i^{Ne}\} ,\{ x_i^{Ge}\} ,\{ x_i^{Xe}\}$, share bins which sample identical regions of the velocity space.  For each bin of the data sets which contains observed events we obtain an inference of $\langle \tilde g\left( v_{\rm min}\right)\rangle_i$ within that bin by taking the value of $\tilde g$ to be constant over the bin width, $\Delta v_{\rm min} = v_{\rm min,2} - v_{\rm min,1}$, so that for bin $i$,
\bea
 x_i =  \langle \tilde g\left( v_{\rm min}\right)\rangle_i {\rm Exp} \sum_j \mathcal{F}_j  \int_{0}^{\infty} & \frac{\left[f_{p}/f_{n} Z_j + (A_j-Z_j)\right]^{2}F_j^{2}\left( E_{R}\right)}{2 \mu_{n\chi}^{2}} {\rm Res} (E_{1,j}, E_{2,j}, E_R) \times \nonumber \\ 
 & {\epsilon} \left(\frac{q_j^{2}\left( E_R\right)}{q_{\rm ref}^{2}}\right)^{n} \left(\frac{q_{\rm ref}^{2}+m_\phi^{2}}{q_j^{2}\left( E_R\right)+m_\phi^{2}}\right)^{2}   dE_{R} \, ,
 \label{eq:momoney}
\eea
where again the index $j$ denotes the target nuclide with number fraction $\mathcal{F}_j$, and $E_{1,j},\ E_{2,j}$ are the $j^{th}$ nuclide's recoil energy bin thresholds which correspond to the original $v_{\rm min}$ bin thresholds $v_{min,1},\ v_{min,2}$.  Eq.~\eqref{eq:momoney} can then be reversed to obtain $\langle \tilde g\left( v_{\rm min}\right)\rangle_i$ for each bin of our synthetic data sets.  Note that whether one recovers $\langle \tilde g_{true}\left( v_{\rm min}\right)\rangle$ or $\langle \tilde g_{infer}\left( v_{\rm min}\right)\rangle$ from Equation~\eqref{eq:momoney} depends on if the DM momentum-dependent interaction physics is included (shown above) or excluded (by setting all momentum transfer dependent terms equal to unity), respectively.}

{Once the full set of data has been converted into a set of measured $\langle\tilde g\left(v_{\rm min}\right)\rangle$, $\{ \langle\tilde g_i^{Ne}\rangle\} $, $\{ \langle\tilde g_i^{Ge}\rangle\} $, $\{ \langle\tilde g_i^{Xe}\rangle\}$, we can compute the ratio of $\tilde g$'s in overlapping $v_{\rm min}$ bins.  We employ the assumption of Poisson random noise in each bin to derive the statistical error associated with the computed ratio, which is then simply the addition in quadrature of the fractional error associated with each individual data bin in the ratio.  This procedure produces a set of three ratio data sets $\{\{ R_i^{Ne/Xe}\} ,\{ R_i^{Ne/Ge}\} ,\{ R_i^{Ge/Xe}\}\}$ and associated uncertainties $\{\{ \sigma_i^{Ne/Xe}\} ,\{ \sigma_i^{Ne/Ge}\} ,\{ \sigma_i^{Ge/Xe}\}\}$.  The expected values of these ratios, $\{\{ \rho_i^{Ne/Xe}\} ,\{ \rho_i^{Ne/Ge}\} ,$ $\{ \rho_i^{Ge/Xe}\}\}$, can be computed easily from Equation~\eqref{eq:gtilderatio} if the $\tilde g_{infer}$ ratio is used.  If the $\tilde g_{true}$ ratio is used, the expected value for all $\mu_i \equiv 1$ .  The $\chi^2$ value of the ratio test for a fixed point in the $m_{X}$, $m_\phi$, $n$ parameter space is then computed directly for the entire data set, along with the accompanying statistical significance.}

{It is important to note that this procedure does introduce a systematic error into the inferred values of $\langle\tilde g\left(v_{\rm min}\right)\rangle$ which is dependent on both the detector response details and the actual distribution of dark matter velocities in the galactic halo $\tilde g_{actual}\left(v_{\rm min}\right)$.  This error arrises directly from the approximation inherent in Eq.~\eqref{eq:momoney} that over the width of a $v_{\rm min}$ bin the function $\tilde g\left(v_{\rm min}\right)$ is constant.  However, the true average of  $\tilde g_{actual}\left( v_{\rm min}\right)$ over this range is given by,
\be
\langle \tilde g_{actual}\left( v_{\rm min}\right)\rangle   =  \frac{1}{\Delta v_{\rm min}}\int_{v_{\rm min,1}}^{v_{\rm min,2}}\tilde g_{actual}(v_{\rm min}^\prime) dv^\prime_{\rm min} \, .
\ee
This leads to the overall systematic error in each determination of $\langle\tilde g\left(v_{\rm min}\right)\rangle$ of,
\be
\sigma_{sys} = \langle\tilde g\left(v_{\rm min}\right)\rangle - \langle \tilde g_{actual}\left( v_{\rm min}\right)\rangle\, .
\ee}

{Empirically, we have found that this additional systematic error is sub-dominant to the statistical error present in our synthetic samples.  Statistically, typical fractional errors in the ratio test are $\sigma_i/R_i \sim O\left( 1-0.1\right)$, which arises from taking the exposure, Exp $= .88\,\rm{ton\times yr}$, along with the LUX limits on the dark matter interaction cross sections.  None of our mock experiments are projected to have more than $\sim 100$ events per bin.  For comparison, the typical fractional error which is introduced systematically by our approach is $\sigma_{sys}/\langle \tilde g_{actual}\left( v_{\rm min}\right)\rangle \sim O\left( 0.01\right)$.  This is easily understood from the form of the SHM velocity distribution function, which is regular and devoid of sharp features.  Thus, for a sufficiently narrow $v_{\rm min}$ bin the function $\tilde g\left( v_{\rm min}\right)$ is linearizable over the bin width and asymptotically $\langle\tilde g\left(v_{\rm min}\right)\rangle - \langle \tilde g_{actual}\left( v_{\rm min}\right)\rangle \rightarrow 0 $.  This is somewhat idealized, however, as finite detector resolution effects will preclude the value of $\sigma_{sys}$ from reaching this asymptotic value when the width of $v_{\rm min}$ bin becomes comparable to the detector resolution.}

\bibliographystyle{JHEP}

\bibliography{nu}

\end{document}